\documentclass[useAMS]{mn2e}
\usepackage{psfig}

\newif\ifAMStwofonts
\AMStwofontstrue

\def\etal{{et al.}}
\def\xte{{\it RXTE}}
\def\asca{{\it ASCA}}
\def\pg{{PG 0804+761}}

\title[X--ray variability of PG 0804+761]
  {The long time-scale X--ray variability of the radio-quiet quasar PG
0804+761} 
\author[I.E. Papadakis et al.]
  {I.E.~Papadakis,$^{1,2}$
  P.~Reig$^{3,1}$ and K.~Nandra$^{4}$ \\
  $^1$ IESL, FORTH-Hellas, 71 110,Heraklion, Crete, Greece \\
  $^2$ Physics Department, University of Crete, 71 003, Heraklion, Crete,
  Greece \\
  $^3$ GACE, Instituto de Ciencias de los Materiales, University of 
  Valencia, P.O. Box 22085, Valencia, Spain \\
  $^4$ Astrophysics Group, Imperial College London, Blackett Laboratory, Prince
  Consort Road, London SW7 2AW, UK } 

\def\LaTeX{L\kern-.36em\raise.3ex\hbox{a}\kern-.15em
    T\kern-.1667em\lower.7ex\hbox{E}\kern-.125emX}

\begin{document}

\label{firstpage}

\maketitle

\begin{abstract}

We present the results from a study of the timing properties and the
energy spectrum of the radio-quiet quasar \pg, based on monitoring \xte
/PCA observations that lasted for a year. This is a systematic study of
the X--ray variations on time scales of weeks/months of the most luminous
radio-quiet quasar studied so far. We detect significant variations in the
$2-10$ keV band of an average amplitude of $\sim 15\%$. The excess
variance of the light curve is smaller than that of Seyfert galaxies,
entirely consistent with the relationship between variability amplitude
and luminosity defined from the Seyfert data alone. The power spectrum of
the \xte\ light curve follows a power-like form of slope $\sim -1$.
However, when we extend the power spectrum estimation at higher
frequencies using archival \asca\ data, we find strong evidence for an
intrinsic steepening to a slope of $\sim -2$ at around $\sim 1\times
10^{-6}$ Hz. This ``break frequency" corresponds to a time scale of $\sim
10$ days. The time-average energy spectrum is well fitted by a $\Gamma\sim
2$ power law model. We also find evidence for an iron line at $\sim 6.4$
keV (rest frame) with $EW\sim 110$ eV, similar to what is observed in
Seyfert galaxies. The flux variations are not associated with any spectral
variation. This is the only major difference that we find when we compare
the variability properties of \pg\ with those of Seyfert galaxies. Our
results support the hypothesis that the same X--ray emission and
variability mechanism operates in both Seyfert galaxies and quasars.

\end{abstract}


%%%%%%%%%%%%%%%%%%%%%%%%%%%%
\section{Introduction}
%%%%%%%%%%%%%%%%%%%%%%%%%%%%

X--ray emission is a universal property of active galactic nuclei (AGN).
The X--rays are known to be variable in all cases where sufficient
signal-to-noise ratio and sampling have been obtained. {\it EXOSAT} first
showed systematically that the X--ray emission of AGN was variable, with
the most convincing demonstration of this being the ``long looks"
(Lawrence \etal\ 1987, McHardy \& Czerny 1987). The data showed no
characteristic time-scales, and the power spectral density function (PSD)  
showed a ``red-noise'' spectrum. Lawrence \& Papadakis (1993) showed that
the AGN PSDs were consistent with a single power law form and an amplitude
which decreases with luminosity. Similar results were obtained by Green,
McHardy \& Lehto (1993). The anticorrelation between variability amplitude
and luminosity was later confirmed by \asca\ (Nandra \etal\ 1997).

Further progress in the study of the longer-term X--ray variability has
been afforded by \xte\ which has shown a flattening of the PSD at lower
frequencies (Edelson \& Nandra 1999, Pounds \etal\ 2001, Uttley, McHardy
\& Papadakis 2002, Markowitz \etal\ 2003). The results so far suggest that
the AGN power spectra are similar (both in shape and fractional rms
amplitude) to those of Galactic black--hole X--ray binaries, like Cyg
X--1, except that characteristic time-scales are ``shifted" to much lower
frequencies. This raises the possibility that the PSDs may scale with the
black hole mass. If this is the case, then the apparent anticorrelation
between variability amplitude and luminosity could be in fact a positive
correlation between luminosity and variability time-scale. If the AGN PSD
shift to lower frequencies as the luminosity increases, the amplitude on a
fixed time-scale would reduce. Therefore, in order to observe large
amplitude variations in high-luminosity sources, we should sample longer
time-scales. Consequently, quasars, being the high-luminosity counterparts
of Seyfert galaxies, should show strong variability on time-scales of
months-years.

In order to investigate the X--ray variability properties of high
luminosity AGN, we performed \xte\ monitoring observations of the X--ray
bright, radio-quiet quasar \pg\ in the period between March 2000 - March
2001. The resulting dataset provides a systematic study of the X--ray
variations of the most luminous radio-quiet quasar studied so far. The
source is one of the brightest radio-quiet quasars which have been
observed by \asca. It has shown significant variations (although of low
amplitude) during a $\sim 1$ day long \asca\ observation. The energy
spectrum was well fitted by a simple power law model with a slope of
$\Gamma \sim 2.2$, while no significant Fe {\it K}-shell emission was
detected (George \etal\ 2000).

In this work we study the long term X--ray variability of \pg\ in terms of
simple statistics like the rms excess variance, and by measuring its 2--10
keV PSD. We compare our results with those from similar studies of
Seyfert galaxies, and with the variability results from the \asca\
observation of the source. We also study the energy spectrum of the
source, and investigate any spectral variations and their relationship to
the flux.

%%%%%%%%%%%%%%%%%%%%%%%%%%%%%%%%
\section{Observations and data reduction}
%%%%%%%%%%%%%%%%%%%%%%%%%%%%%%%%

\xte\ observed \pg\ with the Proportional Counter Array (PCA) every
$\sim 3$ days between March 07, 2000 and March 03, 2001. Individual
observations lasted typically $\sim 1-2$ ksec in all cases. The total
number of observations is 126 and the overall exposure time is 157
ksec. We used FTOOLS v5.2 for the reduction of the PCA data. Data were
collected from the proportional counter unit 2 (PCU2) which was
switched on during almost all of the observations. PCU0 was also
switched on in most cases, however, due to the loss of the propane
layer in May 2000 we do not include data from this unit in our
analysis. Spectra and light curves were extracted from the top Xenon
layer data. We consider only STANDARD-2 data, using ``good time
intervals" determined according to the following criteria: target
elevation $> 10^{\circ}$, pointing offset $<0.02^{\circ}$, and
ELECTRON2 $<0.1$. We calculated background data using the tool
PCABACKEST v3.0 and the new ``CM" models to generate background model
files.

Light curves binned to 16 sec were generated for the source over the
$2-10$, $2-5$, and $5-15$ keV bands, where the PCA is most sensitive
and the background models are best quantified. The light curves were
then re-binned using a 3 day bin size, resulting in evenly spaced
light curves.  The difference between the mid-point of each bin and
the actual observation time of the corresponding points is very small
in all cases (typically only a few percent of the 3 day separation
between successive points in the light curve) and will have negligible
impact on our results.  There are a few ``missing points" in the
resulting light curves which correspond to those observations when
PCU2 was not switched on. The final number of points in each light
curve is 121, with 6 missing points (less than $5\%$ of the total
number of points). We accounted for them using linear interpolation,
adding the appropriate random noise to each point.

All the observations were done during a period when no major gain change
was applied to PCA. However, we extracted light curves individually for
the data prior and after May 13, 2000 (the date of the propane layer loss
of PCU0). We used the appropriate channel boundaries in order to extract
the same energy band light curves for both data sets, following the
relevant information in the ``channel-to-energy" table at the \xte\ web
pages.

%%%%%%%%%%%%%%%%%%%%%%%%%%%%%%%%%%%%%
\section{The $2-10$ keV flux variability}
%%%%%%%%%%%%%%%%%%%%%%%%%%%%%%%%%%%%%

%%%%%%%%%%%%%
\subsection{Basic variability properties}
%%%%%%%%%%%%%

%%%%%%%%%%%%%% Fig. 1 %%%%%%%%%%%%%%%%%%
\begin{figure}
\psfig{figure=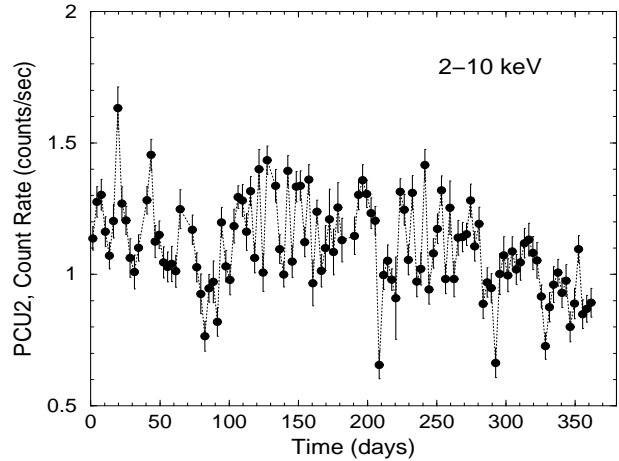,height=7.0truecm,width=9.5truecm,angle=0,%
bbllx=20pt,bblly=10pt,bburx=570pt,bbury=460pt}
\caption[]{The background subtracted 2-10 keV light curve of \pg. Time is
measured in days since the beginning of the monitoring observations. The
source shows significant flux variations on time-scales as short as $\sim$
a few days.}
\end{figure}
%%%%%%%%%%%%%%%%%%%%%%%%%%%%%%%%%%%%%%%%

Fig.~1 shows the $2-10$ keV light curve of the source. Significant
variations can be observed on all sampled time-scales. During most of the
monitoring period the source flux varies around the $\sim 1.2$ counts/sec
level, typically with an amplitude of $\sim 5-30\%$. During the last $2-3$
months of the observation the count rate gradually decreased. First,
following Nandra \etal\ (1997), we parametrized the observed variability
by the ``excess variance" normalized to the mean count rate square
($\sigma^{2}_{rms}$). This quantity is a measure of the integrated
normalised PSD of the source over frequencies between the lowest and
highest sampled frequencies (in our case $1/(363$ days)$\sim 3\times
10^{-8}$Hz, and $1/(3$ days)$\sim 3\times 10^{-6}$ Hz, respectively). We
find $\sigma^{2}_{rms}=0.021$, which implies that the average variability
amplitude of the source is $\sim 15\%$ of the average count rate.

As a first comparison between the variability properties of \pg\ and
Seyfert 1 galaxies, in Fig.~2 we show $\sigma^{2}_{rms}$ plotted against
the $2-10$ keV luminosity for \pg\ and the nine Seyfert galaxies studied
by Markowitz \& Edelson (2001). The normalized excess variance for these
galaxies was estimated from \xte\ $2-10$ keV light curves which had a
duration of 300 days. Consequently, they can be compared directly with the
\pg\ $\sigma^{2}_{rms}$ value estimated in this work. The X--ray
luminosities ($L_{X}$) of the Seyfert galaxies were taken from Markowitz
\& Edelson (2001), while the $2-10$ keV luminosity of \pg\ is estimated as
explained in Section~4.

Fig.~2 shows that the \pg\ measurement agrees very well with the
correlation between $L_{X}$ and $\sigma^{2}_{rms}$ as determined by
the Seyfert data alone. Thus while the variability of \pg\ exhibited
in Fig. 1 is dramatic, the amplitude of variations is smaller than
that of Seyferts, and fits in precesly with our expectations. When we
fit all the data shown in this figure with a power law-model of the
form $\sigma^{2}_{rms}\propto L_{X}^{-\alpha}$, the best fitting slope
value is $\alpha=0.30\pm0.02$. This is consistent with the best
fitting slope found by Markowitz \& Edelson (2001), using only the
Seyfert 1 data.

%%%%%%%%%%%%%% Fig. 2 %%%%%%%%%%%%%%%%%%
\begin{figure}
\psfig{figure=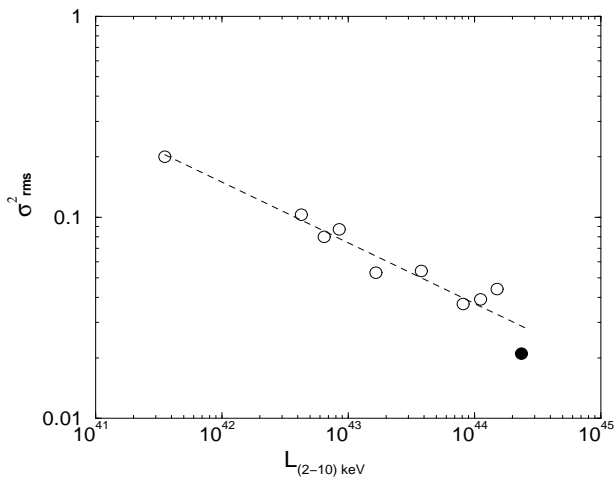,height=6.5truecm,width=7.8truecm,angle=0,%
bbllx=40pt,bblly=30pt,bburx=510pt,bbury=440pt}
\caption[]{The \xte\ excess variance of nine Seyfert galaxies (open circles)
and \pg\ (filled circle) plotted as a function of the source $2-10$ keV
luminosity. The dotted line shows the best-fitting power-law model to the
data.} 
\end{figure}
%%%%%%%%%%%%%%%%%%%%%%%%%%%%%%%%%%%%%%%%

%%%%%%%%%%%%%% Fig. 3 %%%%%%%%%%%%%%%%%%
\begin{figure}
\psfig{figure=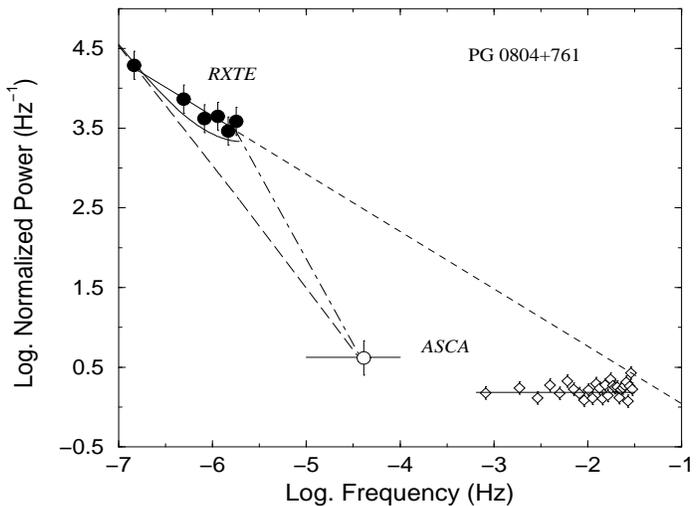,height=7.0truecm,width=8.5truecm,angle=0,%
bbllx=60pt,bblly=35pt,bburx=520pt,bbury=440pt}
\caption[]{The 2-10 keV PSD of \pg\ estimated using the \xte\ monitoring
data presented in this work (filled circles) and the full energy band
($0.5-10$ keV) \asca\ data. The open circle corresponds to the PSD level
using 5760-sec binned \asca\ light curves, and the open diamond points
show the high frequency \asca\ PSD, eastimated using 16-sec binned light
curves. The thin solid line plotted on top of the high frequency \asca\
PSD shows the expected Poisson power level. The best fitting power law
model to the \xte\ PSD is plotted with the tick solid line, while the
dashed line shows its extrapolation to higher frequencies. The dot-dashed
line simply connects the high frequency part of the \xte\ PSD with the
``intermediate-frequency" \asca\ PSD. The long dashed line shows a power
spectrum with slope of $-1.5$ which connects the lowest frequency estimate
of the \xte\ PSD and the \asca\ intermediate-frequency power estimate. The
thin solid line on top of the long dashed line shows the same power
spectrum with aliasing effects and the contridution of the Poisson power
level taken into account (see text for details).}
\end{figure}
%%%%%%%%%%%%%%%%%%%%%%%%%%%%%%%%%%%%%%%%

%%%%%%%%%%%
\subsection{The $2-10$ keV PSD}
%%%%%%%%%%%

As a next step in our study of the X--ray variability of \pg, we
computed the PSD using the interpolated, 3-day binned, $2-10$ keV
light curve.  First, we computed the periodogram (normalized to the
mean count rate square) as in Papadakis \& Lawrence (1993). Then, we
calculated the logarithm of the priodogram estimates, grouped them
into bins (with either 10 or 20 points per bin - see below), and
computed their average value in each bin. The filled circles in Fig.~3
show the \xte\ $2-10$ keV PSD of \pg\ (when using a bin size of 10)
calculated by this method. There are no obvious features or
periodicities, though admittedly the frequency resolution of the PSD
is poor. We have parametrized the spectrum using a power-law model of
the form, $P(\nu) \propto \nu^{-a}$.  The Poisson noise level is very
low. This is immediately obvious from an inspection of the errorbars
in Fig.~1, which are very small when compared to the amplitude of the
observed variations. Indeed the best fitting power law index is
consistent whether or not Poisson noise is accounted for in the fir.
The best fitting power law index for the fit without Poisson noise is
$a=0.72\pm 0.15$, (errors represent the 68\% confidence regions for
the model parameters and were computed using the prescription of
Lampton, Margon \& Bowyer, 1976).  The best fitting power-law model is
shown in Fig.~3 as the thick solid line.  Its slope is much flatter
than the typical value inferred by Lawrence \& Papadakis (1993) from
the analysis of {\it EXOSAT} data of a Seyfert 1 galaxies, or of 
single power law fits to RXTE PSDs od Seyferts (Uttley et al. 2002;
Markowitz et al. 2003). 

Note that the model fitting was performed using the PSD shown in
Fig.~3, i.e. the 10-binned logarithmic periodogram estimates. Although
a bin size of at least 20 is needed for the final PSD estimates to
have the necessary properties for a $\chi^{2}$ model fitting
(Papadakis \& Lawrence, 1993), the small bin size choice was motivated
by the need to increase the frequency resolution of the spectrum,
which is required to estimate reasonable errors for the best-fitting
parameter values. We caution that this may introduce some systematic
error, but note that power-law model fit to the 20-binned logarithmic
periodogram estimates yields similar best-fitting parameter values,
but with a larger uncertainties.

In order to investigate the high frequency PSD and compare it with the
long term PSD, we obtained \asca\ data for \pg\ from the TARTARUS
database. \pg\ was observed with \asca\ for $\sim 89$ ksec on November 11,
1997. In order to maximise the signal-to-noise ratio of the \asca\ data,
we used the full energy band (i.e. $0.5-10$ keV) GIS and SIS light curves.  
This is different to the RXTE data, but we justify our choice by the fact
that in the hard and soft X-ray emission is strongly correlated, both in
general and specifically in this case.  We discuss the impact of this
choice of energy band on our conclusions later.  Because of the earth
occultation gaps, the estimation of the PSD for the \asca\ light curves is
complicated. First, we used 16 sec binned light curves to compute the high
frequency PSD. There are 19 and 13 parts in the GIS and SIS light curves,
respectively, with duration between $\sim 1-3$ ksec and no gaps in them.
For each part we computed the periodogram. Then we combined all 32
periodograms into one file, sorted the periodogram estimates in order of
increasing frequency, calculated their logarithm, grouped them into bins
of 50, and computed their average value at each bin. We caution that in
this way we introduce a systematic bias in the estimation of the source
power spectrum, as many parts of the GIS and SIS light curves overlap in
time. As a result, the source variability will be identical between these
parts, and in principle, the respective periodograms should not be
combined. However, this effect has minimal impact on our results, as the
high frequency PSD, calculated by this method, shows no evidence of the
intrinsic source power spectrum. The PSD, shown in Fig.~3, is flat and
consistent with the expected with the Poisson noise power level (shown
with the thin solid line in the same figure). This result shows that the
high frequency variations are entirely dominated by the Poisson noise
process.

To estimate an ``intermediate-frequency" PSD we used 5760 sec binned GIS
and SIS light curves. The use of the approximate \asca\ orbital period as
the bin size results in an evenly sampled light curves with 16 points in
each one of them. We computed the periodogram of both light curves,
combined them into one file, sorted the periodogram estimates in order of
increasing frequency, subtracted the expected Poisson noise level and
grouped them into one bin of size 16. In Fig.~3, with the open circle
point, we show the resulting PSD estimate. In a way, this point should be
considered as the average normalized source power level over the sampled
frequency range of $10^{-5} - 10^{-4}$ Hz. As mentioned in the previous
paragraph, the use of light curves overlaping in time (like the GIS and
SIS light curves in our case) can introduce a systematic bias in the
estimation of the source PSD. For that reason, we re-computed the $10^{-5}
- 10^{-4}$ Hz source power level using the SIS light curve alone. The
result is very similar to what we obtain when we use both the SIS and GIS
light curves, so any bias has a very small effect on our result.

Note that in Fig.~3 the high frequency PSD includes the Poisson noise
process contribution; it is not possible to measure the intrinsic PSD at
these frequencies, as its amplitude is much smaller than the amplitude of
the Poisson power level.  On the other hand, the ``intermediate-frequency"
power level corresponds to the source power only (i.e. with the
experimental noise contribution subtracted).

Fig.~3 shows clearly that the \xte\ PSD is {\it not} consistent with the
higher frequency PSDs, estimated from the \asca\ data. The low frequency
PSD normalization is much larger that what would be expected from a simple
extrapolation of the \asca\ PSD to lower frequencies. The discrepancy
becomes clearer if we extrapolate the \xte\ PSD best-fitting power-law
model to higher frequencies (dashed line in Fig.~3). If the \xte\ PSD is
representative of the ``true"/intrinsic source power spectrum, and this
spectrum extends all the way up to $\sim 10^{-2}$ Hz, then we should
observe significant variations in the \asca\ light curves even on
time-scales as short as $\sim 100$ sec.

\subsection{Systematic uncertainties}

The main source of systematic uncertainties in the estimation of the \xte\
PSD is the possibility of residual background variations that may have not
been taken into account by the background models and are present in the
$2-10$ keV light curve. However, simultaneous \xte\ and {\it XMM-Newton}
observations of a few Seyfert galaxies during the period covered by our
observations show an excelent aggrement between the \xte\ PCU2 and the
{\it XMM-Newton} count rate (e.g. see Mason \etal\ 2002, in the case of
NGC~4051). Furthermore, the difference between the nornmalization of the
\xte\ and \asca\ PSDs is larger than $\sim 2.5$ orders of magnitude. If we
were to decrease the low-frequency PSD normalisation by that large amount,
the resulting $\sigma^{2}_{rms}$ measurement would be entirely
inconsistent with the relation between $\sigma^{2}_{rms}$ and $L_{x}$, as
determined by the Seyfert 1 data (see Fig.~2).  Finally, the PCA team
estimation of the ``unmodelled variance in (the observed background light
curve) residuals after subtracting the known (background) model
components" is $\sigma \sim 7-9\times 10^{-4}$ (cnts/sec)$^{2}$, for PCU2,
epoch 4 and 5 (see the discussion in C. Markwardt's documentation web
page: lheawww.gsfc.nasa.gov/users/craigm/pca-bkg/bkg-users.html).
Therefore, since the mean value of the \pg\ $2-10$ keV light curve is 1.1
cnts/sec, the residual background contribution to $\sigma^{2}_{rms}$ may
be up to $\sim 7\times 10^{-4}$, or up to $\sim 3.5\%$ of the \pg\
$\sigma^{2}_{rms}$.  This effect cannot account for the difference between
the low and higher frequency PSD shown in Fig.~3.

``Red-noise" and ``aliasing" effects could be another possible source of
uncertainty in the estimation of the \xte\ PSD. Since our results indicate
that there is no significant amount of power at high frequencies, aliasing
effects should not be significant. In order to verify that this is the
case, we considered the case of a power spectrum with a $-1.5$ slope and
an amplitude such that the PSD is consistent with the lowest frequency
estimate of the \xte\ power spectrum and the \asca\ intermediate-frequency
estimate (long dashed line in Fig.~3). Since the observed light curve is
almost evenly sampled, the aliased power at each sampled frequency can be
estimated easily (see e.g. section 7.1.1 in Priestley, 1989). The aliased
power spectrum, including the Poisson noise contribution, is also shown in
Fig.~3 (thin solid line below the \xte\ PSD) and does not fit well the
\xte\ power spectrum. In fact, we find that no power-law model, consistent
with the intermediate-frequency \asca\ power estimate, can fit well the
\xte\ PSD when we include aliasing effects. On the other hand, in the
case of red noise effects, even if the intrinsic PSD continues with a
power law slope of $\sim -1$ down to frequencies as low as $\sim 10^{-8}$
Hz (i.e. a time-scale $\sim 10$ times longer than the length of the \xte\
light curve) the amount of power that will be transferred to the observed
frequency range cannot justify the discrepancy of the 2.5 orders that we
observe between the \xte\ and \asca\ PSDs (e.g. Papadakis \& Lawrence,
1995).

We then investigated the possibility that the \pg\ PSD varies with time
(i.e. the X--ray variability process is non-stationary). For that reason,
we divided the \xte\ light curve in two parts, and computed the
periodogram for each one. The two periodograms look very similar, almost
identical, both in shape and amplitude. This result implies that the PSD
remained roughly constant during the $\sim 1$ year long \xte\ monitoring
observations. Therefore it seems rather unlikely that the large PSD
amplitude difference between the \asca\ and \xte\ observations (which are
separated by $2.5$ years only) is caused by intrinsic PSD normalization
variations.

Another possible source of error is that we have determined the RXTE and
ASCA PSDs in different energy bands. Specifically, the ASCA PSD was
determined with a softer energy band. Even though the soft and hard X-rays
are generally well correlated in AGN, and are in this case, this may be
problematic because the PSD can be energy dependent. For example, Vaughan,
Fabian \& Nandra (2003) found a slope difference of $\sim 0.2-0.3$ in the
case of MCG~-6-30-15. Obviously this could affect the results in detail,
but we note that this difference is not sufficient to explain the
discrepancy between the \xte\ and \asca\ power spectra shown in Fig.~3.
Furthermore, we expect the hard band to show {\it less} integrated power
in the frequency than the soft band, and hence be even more discrepant
with the extrapolation of the RXTE PSD. This is confirmed by analysis of
the excess variance of the ASCA light curve, with the 0.5-2 keV ASCA light
curve $\sigma^{2}_{rms}= 1.1 \times 10^{-3}$ compared to
$\sigma^{2}_{rms}=7.5 \times 10^{-4}$ in the 2-10 keV band. The bandpass
effect would therefore cause an even greater discrepancy between the ASCA
and RXTE PSDs. We have tested this explicitly by calculating the ASCA 2-10
keV PSD. While this is admittedly quite noisy it still lies way below an
extrapolation of the RXTE power.

We conclude that, although existing uncertainties in the background
modeling of the \xte\ background may affect to some extend the shape
and/or amplitude of the \xte\ PSD, the large difference between the \xte\
and \asca\ PSDs shown in Fig.~3 implies the presence of an intrinsic
feature in the PSD. It seems that the \pg\ X--ray PSD follows a $\sim -1$
power-law shape up to $\sim 1\times 10^{-6}$ Hz, and then steepens to a
slope of $\sim -2$ (this ``broken power-law form" is shown with the
dot-dashed line in Fig.~3). This ``characteristic break frequency"
corresponds to a time-scale of $\sim 12$ days.

%%%%%%%%%%%%%%%
\section{The energy spectrum}
%%%%%%%%%%%%%%%

%%%%%%%%%%%%%% Fig. 4 %%%%%%%%%%%%%%%%%%
\begin{figure}
\psfig{figure=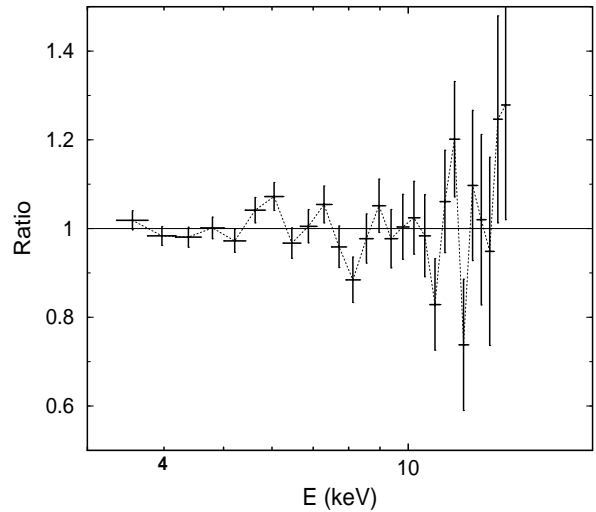,height=7.0truecm,width=7.8truecm,angle=0,%
bbllx=40pt,bblly=30pt,bburx=510pt,bbury=430pt}
\caption[]{Ratio of data/model when a simple ``power law plus absorption  
from neutral material at the redshift of the quasar" model spectrum is     
fitted to the \pg\ spectrum. The presence of an iron line at $\sim 6$ kev
is obvious in the residuals.}
\end{figure}
%%%%%%%%%%%%%%%%%%%%%%%%%%%%%%%%%%%%%%%%

In order to maximize the signal-to-noise ratio, we considered only the
time-averaged spectrum for the \xte\ monitoring observations. We used the
latest response matrix generator in FTOOLS v5.2 in order to generate 12
matrices, one for each month of the monitoring period. The appropriate
detector matrix for the spectrum was created by averaging these matrices.
We used XSPEC v11.2.0 to perform the spectral fitting analysis, and the
errors on the best-fitting parameter values correspond to the 68\% 
confidence limits. Finally, we kept data up to 15 keV, and excluded data
below 3 keV for the spectral analysis due to the uncertainties associated
with the calibration of the softest PCA channels.

First, the spectrum was compared to a simple model of a single power law,
absorbed by neutral material at the redshift of \pg\ ($z=0.1$). This model
provides an adequate description of the data ($\chi^{2}=31.5$, for 24 dof,
prob=0.14). The ratio ``data/model" is shown in Fig.~4. This figure
reveals clearly the presence of an iron line feature at energies $\sim 6$
keV. At energies $> 10$ keV, the residuals appear to be noisy, with no
obvious systematic trend with energy.

The combination of a power law and a Gaussian emission line (plus a
natural material absorption component) results in a better fit to the
spectrum ($\chi^{2}=24.1$, for 22 dof). However, use of the F--test
indicates that the improvement to the goodness of fit is not statistically
significant. The best fitting spectral index is
$\Gamma=2.10^{+0.12}_{-0.10}$, consistent with the results of George
\etal\ (2000) who find $\Gamma = 2.18^{+0.02}_{-0.03}$. The iron line is
not resolved, but the upper limit of 1.36 keV ($\sigma$) provides no
constraint on its origin.  The line's best fitting energy is
$6.46^{+0.31}_{-0.29}$ keV in the rest frame of the quasar, while the
equivalent width is $EW=110 \pm 64$ eV. The best model fitting unabsorbed
$2-10$ keV flux is $1.44\times 10^{-11}$ erg cm$^{-2}$ s$^{-1}$, and the
corresponding X--ray luminosity is $L_{X}=2.4\times 10^{44}$ ergs/s
(assuming $H_{0}=75$ Mpc km$^{-1}$ s, $q_{0}=0.5$).

The best-fitting model together with the observed spectrum and the
residuals plot is shown in Fig.~5. The model describes the overall shape
of the spectrum quite well. There appears to be no systematic deviation of
the data points with energy, which could imply the presence of an extra
component in the spectrum. A possibly origin of the iron line is in
relatively cool optically thick gas, in which case we expect it to be
accompanied by a hard reflection continuum (e.g. Nandra \& Pounds 1994).
We therefore fitted with the XSPEC PEXRAV model (Magdziarz \& Zdziarski
1995), which accounts for this ``Compton reflection'', plus a Gaussian
line. This model does not provide a significantly better fit to the
spectrum ($\chi^{2}_{red}=23.9$, for 21 dof). The best fitting spectral
index, line energy and $EW$ values are almost identical to the values we
find from the previous model, with a best-fitting reflection fraction
value of $R\sim 0.3$. The 90\% confidence upper limit (for one interesting
parameter, i.e. $\Delta\chi^{2}=2.7$) was $R=1.6$, however, which is
consistent with line equivalent width and therefore an origin for the line
in optically thick gas.

%%%%%%%%%%%%%%% Fig. 5 %%%%%%%%%%%%%%%%%
\begin{figure}
\psfig{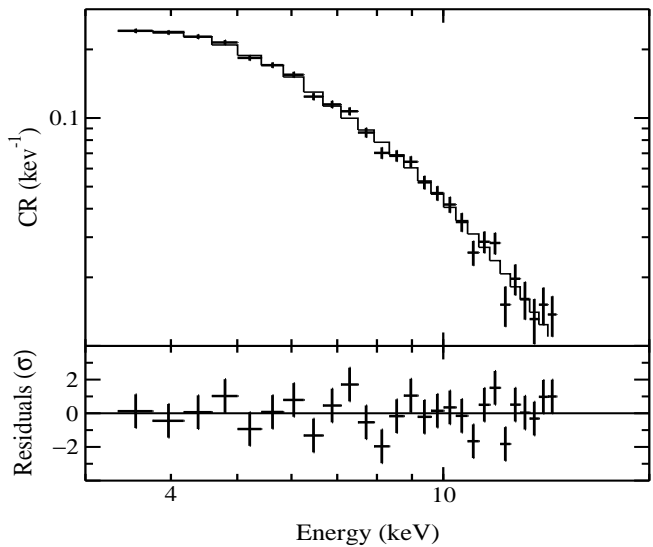}
\caption[]{The \xte\ spectrum of \pg\ fitted with a ``power law plus
Gaussian
and absorption from neutral material at the redshift of the quasar" model.
The lower panel plot shows the model fitting residuals i.e.
(data-model)/$\sigma$, as a function of energy (see text for details).}
\end{figure}
%%%%%%%%%%%%%%%%%%%%%%%%%%%%%%%%%%%%%%%%

%%%%%%%%%%%%%%%%%%%%%%
\subsection{Spectral variability}
%%%%%%%%%%%%%%%%%%%%%%

Apart from the fast, large amplitude intensity variations, Seyfert
galaxies often show flux related spectral variations as well. In order to
investigate the spectral variability properties of the source, we used
\xte\ light curves in the $2-5$ and $5-15$ keV bands, we computed the
hardness ratio, $HR=C_{5-15 keV}/C_{2-5 keV}$, and produced a colour-flux
diagram ($C_{2-5kev}$ and $C_{5-15kev}$ represent the count rate in the
respective energy bands). The two energy bands should be representative of
the primary continuum mainly, as there is no indication of a reflection
component in the energy spectrum, and the iron line emission is not very
strong even in the time-average spectrum, so its contribution in the
$5-15$ keV band during each individual observation should not be
significant. Consequently, the $HR$ values should be sensitive to the
continuum shape variations mainly.

In Fig.~6 we plot the hardness ratio as a function of the $2-10$ keV count
rate, normalised to the mean count rate. The points in this plot cluster
in a well-defined, ``continuous" region rather than forming a scatter
diagram or filling separate ``islands". Therefore, we are certain that we
have observed all the intermediate flux sates between the highest and
lowest flux state of the source in the \xte\ light curves. The open circle
points correspond to the average $HR$ values in flux bins which contain 20
points each.

There is no obvious colour-flux trend in Fig.~6. This becomes clear when
we consider the average $HR$ values (open circle points in Fig.~6) which
do not suggest any flux related variation. When we fit the data with a
power-law function ($HR\propto C^b$) or a linear relationship ($HR=\alpha
+ b\times C$), we find $b \sim 0$. The solid line in Fig.~6 corresponds to
the weighted mean $HR$ value. This line provides a good fit to the data
($\chi^{2}=105.3$ for 114 dof). We conclude that \pg\ shows no spectral
variations. The flux variations correspond to energy spectrum
normalization variations with constant shape.

This is different from what is observed in Seyfert galaxies. For example,
Papadakis \etal\ (2002) study the spectral variability properties of 4
Seyfert galaxies using $\sim 3$ year long, \xte\ light curves. Since we
have used normalized $2-10$ keV values, Fig.~6 can be compared directly
with the plots in Fig.~3 of Papadakis \etal, which show color-flux plots
sensitive to variations of the continuum shape mainly. All Seyfert
galaxies show significant spectral variations which are correlated with the
source flux state: as the flux increases, the spectrum softens (as the
dashed lines in their plots indicate). Typically, $\Delta \Gamma$
variations of the order of $\sim 0.2-0.3$ are observed for maximum
peak-to-peak variations by a factor of $\sim 3-5$. Although the flux
variability amplitude of \pg\ is smaller than the respective amplitude of
the Seyfert galaxies, we should be able to detect similar spectral
variations, if existed, in the case of \pg\ as well.
%%%%%%%%%%%%%%% Fig. 6 %%%%%%%%%%%%%%%%%
\begin{figure}
\psfig{figure=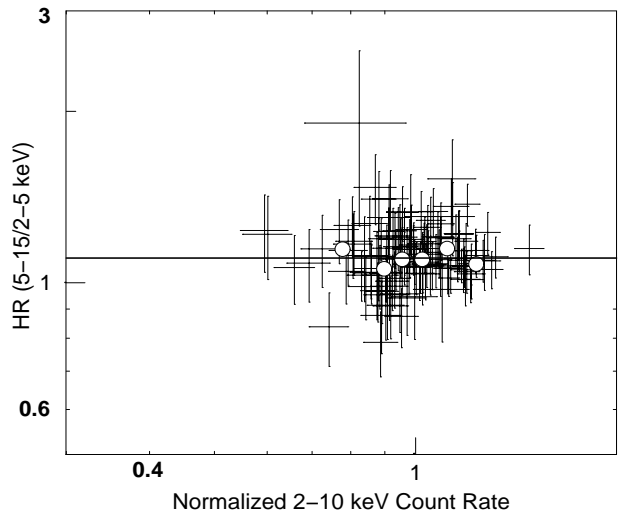,height=7.0truecm,width=8.5truecm,angle=0,%
bbllx=50pt,bblly=40pt,bburx=520pt,bbury=440pt}
\caption[]{Hardness ratio values as a function of the normalized $2-10$ 
keV count rate. Open circles show the average $HR$ values in flux bins    
which include 20 data points. The solid line shows the average $HR$
value.}
\end{figure}
%%%%%%%%%%%%%%%%%%%%%%%%%%%%%%%%%%%%%%%%
%%%%%%%%%%%%%%%%%
\section{Discussion and conclusions}
%%%%%%%%%%%%%%%%%

We have presented a detailed temporal and spectral analysis of the 1 year
\xte\ monitoring observations of \pg. Significant variations are observed
in the $2-10$ keV band. The variability amplitude of the source (as
indicated by the $2-10$ keV $\sigma^{2}_{rms}$) is smaller than the
variability amplitude of Seyfert galaxies, exactly as expected from the
$\sigma^{2}_{rms}$ vs $L_{X}$ relation defined from the Seyfert data
alone. We have presented a power spectrum analysis of the X--ray light
curve, for the case of the most luminous radio-quiet quasar studied so
far. The X--ray PSD follows a power-law like form, with a slope of $\sim
-1$. When combined with higher frequency PSD measurements based on \asca\
data, a strong PSD break (to a slope of $\sim -2$) is implied at $\sim
1\times 10^{-6}$ Hz. The energy spectrum is well fitted by a $\Gamma\sim
2$ power law model plus a Gaussian line at $\sim 6.4$ keV with $EW\sim
110$ eV. Finally, we find that the fast intensity variations are not
associated with significant spectral variations as well. We discuss below
these results in some detail.

\subsection{The PSD break}

The most significant result of the power spectrum analysis is that,
although the \xte\ PSD does not show any characteristic time-scales, its
combination with the \asca\ PSD, implies a significant, strong break at
$\sim 1\times 10^{-6}$ Hz. This break suggests the presence of a
characteristic time-scale in the system of the order of $\sim 12$ days. As
we discussed in Section 3.2, we believe that this break is almost
certainly real, i.e. it represents an intrinsic feature in the PSD of \pg.
Due to the fact that, at present, the \pg\ PSD is poorly defined, we
cannot determine accurately neither the position of the break nor the exact
shape of the PSD. Nevertheless, the available data (shown in Fig.~3) are
consistent with the hypothesis of a PSD steepening from a slope of $\sim
-1$ to $\sim -2$ above the break.

Long term, \xte\ observations have recently revealed the presence of
similar frequency breaks in the PSDs of a few Seyfert galaxies as well
(Edelson \& Nandra, 1999; Pounds \etal\ 2001; Uttley \etal\ 2002,
Markowitz \etal\ 2003).  Both the excellent agreement of the \pg\
$\sigma^{2}_{rms}/L_{X}$ measurement with the $\sigma^{2}_{rms}$ vs
$L_{X}$ relationship defined for Seyfert galaxies, and the similarity
between the shape of its overall PSD with the power spectra of other
Seyfert galaxies support the idea that that the same variability process
operates in both Seyfert galaxies and quasars.

It is interesting to compare the ratio of the break time-scales and black
hole mass between \pg\ and the Seyfert galaxies studied by Markowitz
\etal\ (2003). These authors find that the $2-10$ keV PSD of three Seyfert
1 galaxies, namely NGC~3783, NGC~3516 and NGC~4151, shows a steepening
from a $-1$ to a $-2$ slope at $4\times 10^{-6}$ Hz, $2\times 10^{-6}$Hz,
and $1.3\times 10^{-6}$ Hz, respectively. The black hole mass of these
three objects is $1.1\times 10^{7}$ M$_{\odot}$ (Kaspi \etal\ 2000),
$1.7\times 10^{7}$ M$_{\odot}$ (Onken \etal\ 2003) and $1.2\times 10^{7}$
M$_{\odot}$ (Kaspi \etal\ 2000), respectively. According to Kaspi \etal\
(2000), the black hole mass of \pg\ is $\sim 1.6\times 10^{8}$
M$_{\odot}$. Assuming that the \pg\ PSD shows a break frequency at
$1\times 10^{-6}$, then, on average, we find that
$\nu_{bf}$(Seyferts)$/\nu_{bf}$(\pg)$\sim 2.5$ while
$M_{BH}$(\pg)$/M_{BH}($Seyferts$)\sim 12$. If we take account of the
uncertainties associate with the determination of both $\nu_{bf}$ and
$M_{BH}$, we believe that this result is consistent with the hypothesis
that the characteristic time-scales scale linearly with the black hole
mass in Seyfert galaxies and quasars.

We would like to stress the fact that the detection of a break frequency
in the PSD of \pg\ is only suggestive at the moment. More data are needed
in order to estimate better the overall PSD for \pg\, and measure
accurately $\nu_{bf}$. However, we think it is rather remarkable that the
present results are consistent with the hypothesis of a linear relation
between break time-scale and $M_{BH}$ in Seyferts and quasars. This result
becomes even more remarkable if we also consider the similarity in the PSD
shape and the linear mass-time-scale linear relation between Seyfert and
Galactic X--ray black hole binaries suggested by Edelson \& Nandra (1999),
Uttley \etal\ (2002) and Markowitz \etal\ (2003). The similarity of the
PSD shape implies that the same variability mechanism operates in
accreting systems with an enormous black hole mass range
($10-10^{8}$M$_{\odot}$). Furthermore, if the characteristic time-scale
increases linearly with the black hole mass, then the main factor which
determines the variability properties of the accreting black hole systems
could simply be the size of the X--ray emitting source. As the black hole
mass increases, the source size should increase as well. As a result time
scales should also increase, and it would take longer time for a source to
exhibit a fixed variability amplitude.

\subsection{The energy spectrum and the lack of spectral variability}

The spectral analysis results of the \xte\ data presented in this work
further support the idea that the same physical mechanism operates in both
Seyfert galaxies and in their higher luminosity counterparts, the
radio-quiet quasars. The best-fitting power law index of \pg\ is
consistent with the average spectral slope of Seyfert 1 galaxies, $\Gamma
\sim 1.95$ (Nandra \& Pounds, 1994). The line energy suggests emission by
fluorescence in near-neutral material. The equivalent width of the line is
also consistent with the average value found by Nandra \& Pounds (1994),
from a fit of narrow Gaussians to {\it GINGA} data of Seyfert 1 galaxies.

The lack of spectral variability is the only major difference that we find
between the X--ray properties of \pg\ and Seyfert galaxies. The $2-10$ keV
flux variations in Seyferts are almost always associated with correlated
spectral variations (e.g. Papadakis \etal, 2002, and references therein).
Recent work on the spectral variability of Seyfert galaxies suggest that
the spectral variations is the result of summing two components with
different but intrinsically non-varying spectral shapes(e.g. Taylor,
Uttley \& McHardy, 2003, and references therein). In the case of \pg\ we
find that the $2-10$ keV variability in occurs with a constant spectral
slope. It seems then that the X--ray emission of \pg\ is dominated by a
single power law component of variable normalization. Furthermore, this
result suggests intrinsic luminosity variations in the absence of
associated variations of the properties of the ``hot corona"  (assuming
the currently favoured thermal Comptonization models where hot electrons
produce the X--rays by inverse-Compton upscattering of soft, i.e.
optical/UV, photons;  Sunyaev \& Titarchuk 1980; Haardt \& Maraschi 1991).
However, these models also predict spectral slope variations. An increase
of the soft input photon flux will result in the reduction of the corona
temperature, a steepening of the intrinsic spectrum, and hence an increase
in the $2-10$ keV flux. This response of the X--ray spectrum to the soft
input photon changes has already been observed in Seyfert galaxies (e.g.
NGC~7469, Nandra \etal\ 2000; NGC~5548, Petrucci \etal\ 2000). One
possible scenario that can explain the absence of spectral variations in
\pg\ is described by Haardt, Maraschi \& Ghisellini (1997) in the case of
a pair dominated Corona. If pairs dominate the scattering opacity of the
hot corona, then during flux variations within a factor of 2 or so (the
case of \pg) the spectral index is expected be remain essentially
constant.

The corona is pair dominated in the case when the ``compactness"
parameter, $l_{c}$, is larger than $\sim 10$. In the case of the two-phase
disc-corona model (e.g. Haardt \& Maraschi, 1991), $lc\propto (L_{C}/R)$,
where $L_{C}$ is the overall X--ray luminosity and $R$ is the typical
size-scale of the X--ray emitting region. According to the variability
results mentioned in the previous section, if the main parameter that
determines the variability properties of AGN is the black hole mass
through the relation $R\propto M_{BH}$ (see discussion in section 5.1) and
if $L_{C}\propto M_{BH}$ then we would expect the ratio $L_{C}/R$ to
remain constant in Seyferts and quasars. The fact that the corona could be
pair dominated in \pg\ but not in Seyfert galaxies (where spectral index
variations are observed)  could imply that $L_{2-10keV}$ is not an
accurate indicator of the total X--ray luminosity in \pg\ (for example the
cut-off energy in the X--ray spectrum could be higher than in Seyfert
galaxies). However, in this case, the agreement between \pg\ and Seyferts
shown in Fig.~2 should be accidental. Another possibility is that the
geometry of the X--ray emitting region is different in \pg. Instead of a
single smooth corona, the X--ray region may consist of numerous small
regions above the disc (patchy corona). In this case $l_{c}$ could be
increased, hence leading in a pair-dominated corona. Even if this is the
case, if the absence of intrinsic spectral variations is due to a pair
dominated corona, the energy spectrum of the source should extend up to at
least $\sim 500$ keV, in contrast to the X--ray spectra of Seyfert
galaxies which are thought to show a high energy cut-off at a few hundred
keV.

\section*{Acknowledgments}

PR acknowledges partial support from the European Union via the Training
and Mobility of Researchers Network Grant ERBFMRX/CT98/0195 and from the
Generalitat Valenciana via the Programme Ayudas para las acciones de apoyo
a la investigaci\'on. PR is a researcher of the programme {\it Ram\'on y
Cajal} funded by the University of Valencia and the Spanish Ministry of
Science and Technology. This research has made use of the TARTARUS
database, which is supported by Jane Turner and Kirpal Nandra under NASA
grants NAG5-7385 and NAG5-7067. We would like to thank an anonymous
referee for a fast and helpful report.

\end{document}